\documentclass[%
reprint,
superscriptaddress,
showpacs,
amsmath,amssymb,
aps,
pre,
]{revtex4-1}

\usepackage{graphicx}
\usepackage{dcolumn}
\usepackage{bm}

\usepackage{ulem} 
\usepackage{color}

\newcommand{\removed}[1]{}

\begin{document}

	\title{Cooling by Heating:\\Restoration of the Third Law of Thermodynamics}
	
	\author{V. B. S{\o}rdal}
	\email{v.b.sordal@fys.uio.no}
	\affiliation{Department of Physics, University of Oslo, PO Box 1048 Blinderm, 0316 Oslo, Norway}
	\author{J. Bergli}%
	\affiliation{Department of Physics, University of Oslo, PO Box 1048 Blinderm, 0316 Oslo, Norway}
	\author{Y. M. Galperin}
	\affiliation{Department of Physics, University of Oslo, PO Box 1048 Blinderm, 0316 Oslo, Norway}
	\affiliation{Ioffe Institute, 26 Politekhnicheskaya, St Petersburg 194021, Russian Federation}

	\date{\today}
	
	\begin{abstract}
    We have made a simple and natural modification of a recent quantum  refrigerator model presented by Cleuren \textit{et al.} in Phys. Rev, Lett. \textbf{108}, 120603 (2012). The original model consist of two metal leads acting as heat baths, and a set of quantum dots that allow for electron transport between the baths. It was shown to violate the dynamic third law of thermodynamics (the unattainability principle, which states that cooling to absolute zero in finite time is impossible), but by taking into consideration the finite energy level spacing in metals we restore the third law, while keeping all of the original model's thermodynamic properties intact.
	\end{abstract}
	
	\pacs{05.70.Ln}
	\maketitle
	
	
\section{\label{Introduction}Introduction}
Quantum refrigerators are solid-state devices with huge potential benefits in technology. With no moving parts and of microscopic size, they could easily be integrated into existing technology, such as cellphones and computers, to enhance their performance by utilizing the waste heat energy they produce. As always, the technological frontier is supported by a backbone of theoretical framework, which in recent years have seen many advancements, 
see, e.g.,~\cite{PhysRevE.91.050102,PhysRevB.85.075412,PhysRevLett.110.256801,PhysRevE.64.056130,PhysRevLett.105.130401,PhysRevLett.108.070604}. In addition to the technological possibilities they present, quantum refrigerators are excellent tools for providing insight into the unique features of open quantum systems. For a review of stochastic thermodynamics and the formalism used to treat quantum refrigerators see, e.g.,~\cite{0034-4885-75-12-126001} and \cite{doi:10.1146/annurev-physchem-040513-103724}. 

The quantum absorption refrigerator is a version of these general machines, based on producing a steady-state heat-flow from a cold to a hot reservoir, driven by absorption from an external heat reservoir.
A key tool to understand the operation of these refrigerators, when approaching the limiting temperature of absolute zero, is the laws of thermodynamics. In this article we have studied one such device that appeared to violate the dynamic version of the third law of thermodynamics (the unattainability principle), which states that one can not cool a system to absolute zero in a finite amount of time. A recent publication by Cleuren \textit{et al.}~\cite{PhysRevLett.108.120603} presented a novel model based on two electronic baths coupled together via a system of quantum dots and driven by an external photon source. The article generated some controversy due to its apparent violation of the unattainability principle, and several authors \cite{PhysRevLett.109.248901,PhysRevLett.109.248902,PhysRevLett.109.248903,PhysRevLett.112.048901} proposed explanations for this violation. However, we find that the discussion was without conclusion, and we will discuss this later in the article.

We will begin by giving a brief presentation of the quantum refrigerator model, as introduced by Cleuren \textit{et al.}~\cite{PhysRevLett.108.120603}, and its thermodynamic properties. Then we will summarize and comment on the discussion that followed. Finally we will present a simple modification, based only on the fact that the energy levels of metals are discrete when treated quantum mechanically, which becomes important at temperatures $ T\lesssim\Delta $ where $ \Delta $ is the level spacing. (We measure temperature in energy units putting the Boltzmann constant $k_B=1$).
Our modification upholds the third law, while it simultaneously reproduces the results from the original model down the the limit of $ T\sim\Delta $. 
In essence, we want to make the point that the unattainability principle is only valid when applied to a quantum description of a system.

\subsection{\label{model}The Model}
\begin{figure}[b]
\centering
\includegraphics[width=0.8\linewidth]{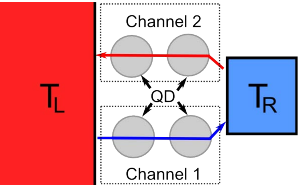}
\caption{A schematic of the model shown in real-space. A small piece of metal with temperature $ T_R $ is coupled to a larger piece with temperature $ T_L>T_R $. Four quantum dots form two channels for electron transport between the metals. The arrows indicate the desired direction of the net particle current to achieve cooling of the right metal lead. The distance between the two channels is too large for any Coulomb interaction to take place between them.}
\label{fig:real_model}
\end{figure}
\begin{figure}[ht]
	\centering
	\includegraphics[width=\linewidth]{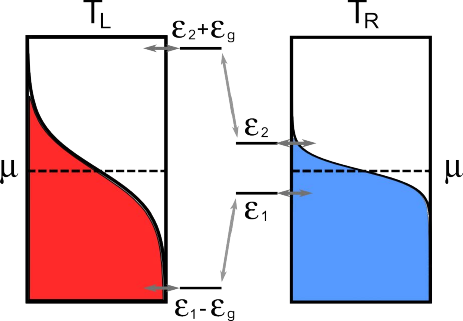}
	\caption{A hot metal lead ($ T_L $) is coupled to a cold one ($ T_R $) via two spatially separated pairs of quantum dots, which form  two channels for electron transport between the leads. We consider the case where $ \mu_L=\mu_R=\mu $, and the energy levels of the quantum dots are symmetric about the chemical potential ($ \epsilon_2-\mu=\mu-\epsilon_1 \to \epsilon_1=-\epsilon_2$). Schematic adapted from \cite{PhysRevLett.108.120603}.}
	\label{fig:model}
\end{figure}
The quantum refrigerator model proposed in \cite{PhysRevLett.108.120603} is shown schematically in real-space in Fig.~\ref{fig:real_model} and in energy-space in Fig.~\ref{fig:model}. Here we will briefly explain its operating protocol. It consists of two metal leads and four quantum dots; the large and hot lead with temperature $ T_L $ is coupled to the small, cold lead with temperature $ T_R $, via the set of quantum dots. We assume that each quantum dot is highly confined, and is thus associated with a single energy level, since the other levels are far outside the energy-range of the system. These four levels are marked in Fig.~\ref{fig:model}. The quantum dots form two channels, as illustrated in Fig.~\ref{fig:real_model}, where the energy levels $ \epsilon_2 $ ($ \epsilon_1 $) and $ \epsilon_2+\epsilon_g $ ($ \epsilon_1-\epsilon_g $) are coupled together in channel 2 (channel 1). The two channels are spatially separated, therefore we can safely ignore any Coulomb interaction between the electrons in channel 1 and 2. The basic idea is to move cold electrons (i.e, with energy less than $ \mu $) from the hot lead into the cold lead via channel 1, while simultaneously move hot electrons (energy greater than the chemical potential $ \mu $) from the cold lead to the hot lead via channel 2. This transport of electrons will thus cool the right lead by injecting cold and extracting hot electrons. Naturally the transport will also heat up the left lead, but since we assume that it is a large piece of metal with a high heat capacity, the heat absorbed will not result in a measurable change in $ T_L $. We can obtain the desired particle flow direction by coupling the quantum dot system to a bosonic bath which induces transitions between the quantum dots of each pair, i.e., between $ \epsilon_1 $ and $ \epsilon_1-\epsilon_g $ in channel 1, and between $ \epsilon_2 $ and $ \epsilon_2+\epsilon_g $ in channel 2. The bosonic bath can be photons from an external source, and/or phonons from the device. In this discussion we will consider it to be a photon bath with temperature $ T_S $. In Ref.~\cite{PhysRevLett.108.120603} the photon bath is taken to be the Sun with a temperature $ T_S\simeq6000$~K, and we will follow this in the sense that we will assume that it is the largest energy scale in the system. In any case the transition rates between the quantum dots are proportional to the probability of finding a boson with energy equal to the energy-difference between the two quantum dot levels, which is given by the Planck distribution $ n(E) $. The rates are thus given by
\begin{equation} \label{eq:rates1}
 k_{\uparrow}^{\epsilon_g} =\frac{\Gamma_s}{e^{\epsilon_g/T_S}-1}, \quad  k_{\downarrow}^{\epsilon_g}=\frac{\Gamma_s}{1-e^{-\epsilon_g/T_S}}.
\end{equation} 
Here $ k^{\uparrow} $ and $ k^{\downarrow} $ are the rates for upwards and downwards transitions in energy, respectively. The difference between them is that $ k^{\downarrow} $ contains an additional term for spontaneous emission.

The transition rate for electron transfer from the metal to an empty quantum dot level is proportional to the probability of finding an electron in the same energy level in the metal, which is given by the Fermi-Dirac distribution $ f(E) $. For the inverse transition to take place there has to be an available energy level in the metal, which has a probability proportional to $ 1-f(E) $. Thus the transition rates between quantum dot and metal are
\begin{equation} \label{eq:rates2}
 k^{E}_{l\to d} =\frac{\Gamma}{e^{(E-\mu)/T}+1}, \quad
 k^{E}_{d\to l}=\frac{\Gamma}{e^{(\mu-E)/T}+1}.
\end{equation}
For transitions involving the right lead the temperature $ T=T_R $, while for the left lead $ T=T_L $. Notice that in general $ \Gamma\neq\Gamma_s $. These are the constants that set the timescale of the transitions and depends on the specific details of the device.

As in Ref.~\cite{PhysRevLett.108.120603}, we will considering the strongly coupled case where the energies of the quantum dots are symmetric about the chemical potential ($ \epsilon_2-\mu=\mu-\epsilon_1$). We can therefore choose to measure all energies relative to $ \mu=0 $, and combine the two parameters $ \epsilon_2=-\epsilon_1=\epsilon $. 

We can now introduce three distinct occupation probabilities per channel. Since the two quantum dots in the same channel are close to each other in space we assume that the Coulomb repulsion between electrons prevents simultaneous occupation of the right and left quantum dot. For channel 1 we then have the probabilities $ P^{(1)}_L,P^{(1)}_R$, and $P^{(1)}_0$, which represent the probability of finding an electron in the left quantum dot with energy $ -(\epsilon+\epsilon_g) $, in the left quantum dot with energy $ -\epsilon $, and in neither quantum dot, respectively.
A master equation describing the time-evolution of the occupation probabilities in channel 1 can thus be formulated:
\begin{equation}
\label{master_eq}
\dot{\mathbf{P}}^{(1)}=\hat{M}^{(1)} \mathbf{P}^{(1)}\, , \quad \mathbf{P}^{(1)} \equiv \begin{bmatrix}
P^{(1)}_0 \\
P^{(1)}_L \\
P^{(1)}_R 
\end{bmatrix},
\end{equation}
where the transition matrix $ M^{(1)} $ is given by
\begin{equation}
\nonumber
M^{(1)}
\! \!  \!= \! \! \!
\begin{bmatrix}\! 
-k_{l\to d}^{-(\epsilon+\epsilon_g)}-k_{l\to d}^{-\epsilon} & k_{d\to l}^{-(\epsilon+\epsilon_g)} & k_{d\to l}^{-\epsilon} \\\\
k_{l\to d}^{-(\epsilon+\epsilon_g)} & -k_{d\to l}^{-(\epsilon+\epsilon_g)}-k_{\uparrow}^{\epsilon_g} & k_{\downarrow}^{\epsilon_g} \\\\
k_{l\to d}^{-\epsilon} & k_{\uparrow}^{\epsilon_g} & -k_{d\to l}^{-\epsilon}-k_{\downarrow}^{\epsilon_g} \! \!
\end{bmatrix} \! \! .
\end{equation} 

We are interested in the steady state of the system, where the probabilities does not change as a function of time. To find this state we set $ \dot{\mathbf{P}}^{(1)}=\mathbf{0} $, and solve Eq.~(\ref{master_eq}). By doing this we obtain the steady state probability vector $\mathbf{P}^{(1)}(\epsilon,\epsilon_g,T_R,T_L $) where we consider $ \Gamma,\Gamma_s $ and $ T_S $ as constants. A similar procedure gives us the steady-state probability vector for the channel 2 as well.

The particle current between the right dot in the lower level and the cold lead can be written as
\begin{equation}
J^{(1)}= P^{(1)}_Rk_{d\to l}^{-\epsilon} - P^{(1)}_0k_{l\to d}^{-\epsilon} \, ,
\end{equation}
and the current through the upper level is
\begin{equation}
J^{(2)}= P^{(2)}_Rk_{d\to l}^{\epsilon} - P^{(2)}_0k_{l\to d}^{\epsilon} \, .
\end{equation} 

The cooling power, i.e., the heat transported \textit{out of} the right lead per unit time, can now be defined as:
\begin{equation}
\dot{Q}_R=(-\epsilon-\mu)(-J^{(1)}) + (\epsilon-\mu)(-J^{(2)}).
\end{equation}
Since the energy levels are symmetric about $ \mu $, we can set $ \mu=0 $, and we obtain the cooling power for the refrigerator model.
\begin{equation}
\dot{Q}_R=\epsilon(J^{(1)}-J^{(2)}).
\end{equation}
Optimized cooling is attained by varying $ \epsilon(T_R) $ and $ \epsilon_g(T_R) $ as a function of $ T_R $ (when $ T_S $ and $ T_L $ are kept constant). It can be shown (see Ref.~\cite{PhysRevLett.108.120603} for details) that the cooling power in the limit of low $ T_R $ is given by
\begin{equation}
\lim\limits_{T_R\to 0}\dot{Q}_R\propto  T_R .
\label{eq:Q_lowT}
\end{equation}

When working at an energy-scale where $\epsilon_g\ll T_S $ we have $ k^{\uparrow}\simeq k^{\downarrow} $. In this situation we can get a better understanding of the system and when cooling will occur by considering the transitions in channel 2. There the energy levels are situated above $ \mu $, and we have 
\[ 0< f(E)< 1/2, \quad  1/2<1-f(E)<1. \]
Therefore the rate from lead to dot will always be less than the rate from dot to lead, $ k_{l\to d}^{E}<\ k_{d\to l}^{E} $, for a given energy $ E $. The requirement for cooling to take place in this situation is that $ f(\epsilon+\epsilon_g)<f(\epsilon) $, i.e., we require $ (\epsilon+\epsilon_g)/T_L>\epsilon/T_R $. We then have
\begin{equation}
\left. \begin{array}{l}
 k^{\epsilon+\epsilon_g}_{d\to l} > k^{\epsilon}_{d\to l} \\
  k^{\epsilon+\epsilon_g}_{l\to d} < k^{\epsilon}_{l\to d} \\
   k_{l\to d}^{E}<\ k_{d\to l}^{E}
   \end{array}\right\} \Rightarrow 
k_{d\to l}^{\epsilon+\epsilon_g}>k_{d\to l}^{\epsilon}>k_{l\to d}^{\epsilon}>k_{l\to d}^{\epsilon+\epsilon_g} \, .
\label{rate_inequalities}
\end{equation}
%
When $ k^{\uparrow}\simeq k^{\downarrow} $ we know that the occupation probability $ P_L^{(2)}\simeq P_R^{(2)} = P $, and thus $ P_0^{(2)}=(1-2P) $. Using the inequalities shown in Eq.~(\ref{rate_inequalities}) we now consider two different states of the system. First assume that there is an electron in the quantum dot system; it can either exit into the left lead or the right lead, where the currents are $k^{\epsilon+\epsilon_g}_{d\to l}P_L^{(2)} $ and $ k^{\epsilon}_{d\to l}P_R^{(2)} $, respectively. The difference is
\[ P\big(k^{\epsilon+\epsilon_g}_{d\to l}- k^{\epsilon}_{d\to l} \big)>0 \]
that tells us it is more likely for the electron to exit into the left lead. Next we assume the quantum dot system is unoccupied; an electron can enter from the left lead or the right lead, with currents $k^{\epsilon+\epsilon_g}_{l\to d}P_0^{(2)} $ and $k^{\epsilon}_{l\to d}P_0^{(2)} $, respectively. The difference is now
\[ (1-2P)\big(k^{\epsilon+\epsilon_g}_{l\to d}-k^{\epsilon}_{l\to d}\big)<0 \]
indicating that it is more likely that an electron enters from the right lead. Above the chemical potential, electrons entering from the right lead and exiting into the left lead corresponds to a net cooling of the right lead, which is our desired effect. A similar analysis can be done for channel 1, where the corresponding result of net transport from the left to the right lead is obtained.

\subsection{\label{unatt}The Unattainability Principle}
The unattainability principle states that one cannot cool a system to absolute zero in a finite amount of time~\cite{PhysRevE.85.061126}. A system with heat capacity $ C_V=dQ/dT $ and cooling power $ \dot{Q}\equiv dQ/dt $ has a cooling rate given by
\begin{equation}
\frac{dT}{dt}=\frac{\dot{Q}}{C_V}.
\end{equation}
If we assume that $ C_V $ and $ \dot{Q} $ scale with temperature to the power of $ \kappa $ and $ \lambda $, respectively, we have 
\begin{equation}
\frac{dT}{dt}\propto T^{\lambda-\kappa}.
\label{eq:unattainability}
\end{equation}
For $ \alpha\equiv\lambda-\kappa < 1 $ the unattainability principle is violated~\cite{PhysRevLett.109.248901}, and cooling to absolute zero is possible in finite time. By inspecting Eq.~(\ref{eq:Q_lowT}) we find that $ \lambda=1 $. The heat capacity of the metal lead as $ T_R\to 0 $ is dominated by the electronic heat capacity, which is  proportional to the temperature $ C_V\propto T_R $ (see chapter 7.H.2 of Ref.~\cite{reichl1980modern} or an equivalent textbook), and therefore $ \kappa=1 $. The end result is that $ \dot{T}_R\propto T^0 $, in violation of the unattainability principle.
\subsection{\label{comments}Comments}
Levy \textit{et al.}~\cite{PhysRevLett.109.248901} were the first to point out that because the refrigerator presented in \cite{PhysRevLett.108.120603} has a cooling power of $ \dot{Q}\propto T_R $ and a heat capacity of $ C_V\propto T_R $ in the limit of $ T_R\to 0 ~\mathrm{K} $, its cooling rate is given by
\begin{equation}
\frac{dT(t)}{dt}=\frac{\dot{Q}}{C_V}\propto T_R^0=\text{const} .
\end{equation}
That enables cooling to absolute zero in a finite amount of time. In the original model proposed by Cleuren \textit{et al.} the quantum dot system consisted of only two quantum dots, with the levels $ \epsilon_1 $ ($ \epsilon_1-\epsilon_g $) and $ \epsilon_2 $ ($ \epsilon_2+\epsilon_g $) being two adjacent levels within the right (left) quantum dot. Levy \textit{et al.} suggest that the violation of the third law may be due to the neglect of internal transitions within a single dot. This suggestion was refuted by Cleuren \textit{et al.}~\cite{PhysRevLett.109.248902} who stated that  the model could also be constructed using two pairs of spatially separated quantum dots, as we have done here. Their own explanation for the violation was that the quantum master equation they utilized does not take into account coherent effects, and the broadening of the linewidth of the quantum dot energy levels was ignored. Both of these effects becomes important in the low-temperature limit.

Allahverdyan \textit{et al.}~\cite{PhysRevLett.109.248903} suggested that the violation occurs since the weak-coupling master equation used by Cleuren \textit{et al.} is limited at low temperatures. They state one can justify taking the limit, $ T_R\to0 $, for such an equation only while simultaneously reducing the coupling between the quantum dot system and heat reservoirs, $ \gamma \to 0 $. Concrete analysis of the low-temperature behavior of the cooling power is not given.

Finally, Entin-Wohlman \textit{et al.}~\cite{PhysRevLett.112.048901} considered a simplified version of the original model, where only a single channel contribute to the electron transfer. They assume that boson-assisted hopping is the dominant form of electronic transport \cite{PhysRevB.85.075412} (an assumption we will also make later in the article). If we remove channel 1 from our model and only consider channel 2, we obtain the same system as considered in \cite{PhysRevB.85.075412}. Using Fermi's golden rule they find that the heat current is exponentially small for $ \epsilon_2-\mu \gg T_R $. They go on to state that the violation of the third law comes from allowing the levels $ \epsilon_1 $ and $ \epsilon_2 $ to  approach the chemical potential linearly as a function of temperature, and claim that this is unnecessary and complicates the setup. In our opinion, the linear temperature dependence of the energy levels $ \epsilon_1 $ and $ \epsilon_2 $ in the quantum dots coupled to the cold lead is an essential feature --  it arises from the optimization of the cooling power suggested in Ref.~\cite{PhysRevLett.108.120603}, but not implemented in Ref.~\cite{PhysRevLett.112.048901}.

\section{\label{discretization}Discretization of the Model}

One of the assumptions of the model proposed is that there is a continuous spectrum of energy states in the metal leads. Thus the electrons are transferred elastically between the quantum dots and the metals. We will now introduce a simple discretized modification of the original model, and show that the unattainability principle will then be restored.
In our model, we will assume an even spacing between the energy levels.
\begin{figure}[t]
\centering
\includegraphics[width=0.7\linewidth]{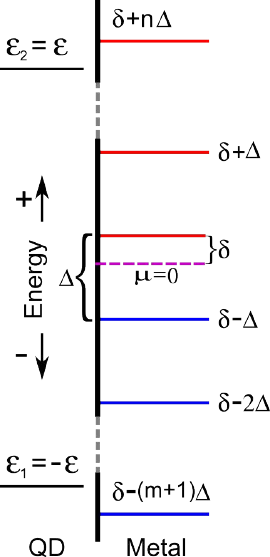}
\caption{The continuous states of the metal are replaced by a discrete spectrum with a constant energy-spacing $ \Delta $. The asymmetry between states above and below $ \mu $ is modeled by the parameter $ \delta $. For $ \delta=\Delta/2 $ the chemical potential lies exactly in the middle of two energy-levels. The $ j $th [$ i $th] level below [above] $ \mu $ is given by $ \epsilon_j=\delta-j\Delta $ [$ \epsilon_i=\delta+(i-1)\Delta $].}
\label{fig:d_model}
\end{figure}
We also introduce the parameter $ \delta $ to quantify the asymmetry about the chemical potential $ \mu $, see Fig.~\ref{fig:d_model}. If $ \delta=\Delta/2 $ the energy levels are symmetrically distributed about $ \mu $. 
As long as the quantum dot and metal energy levels do not exactly overlap, the transitions are now inelastic and require absorption/emission of phonons.

\subsection{Cooling Power}
We can set up a master equation for the dynamics in channel 1, as in Eq.~(\ref{master_eq}), but now for the discrete system. The rate-matrix is almost identical, but since we allow for phonon-assisted transitions, the rates between the quantum dots and the discrete levels of the right lead are given by a sum of all possible emission and absorption transitions.
We will use $ \epsilon_{n} $/$ \epsilon_{m}$ to denote the $ n $th/$ m $th level in the metal lead, above/below the quantum dot level $ \epsilon_1 $. We also introduce $ \omega_n=\epsilon_{n}-\epsilon_1 $ and $ \omega_m=\epsilon_1-\epsilon_{m} $ to represent the phonon frequencies associated with transitions between these levels. For transitions from the lead to the dot, $ \epsilon_n $ and $ \epsilon_m $ are the energies associated with emission and absorption processes, respectively, while for dot-to-lead transitions the association is opposite. The  matrix elements changes from $ k^{\epsilon_1}_{d\to l}\to k^{d,\epsilon_1}_{d\to l} $ and $  k^{\epsilon_1}_{l\to d}\to  k^{d,\epsilon_1}_{l\to d} $, where we use the superscript $ d $ to indicate that it is the transition rate for the discrete model. These rates are then sums of all possible emission and absorption processes, and can be written as
\begin{align}
 k^{d,\epsilon_1}_{d\to l}&=\overbrace{\sum\limits_{m}k^{\epsilon_m}_{d\to l}}^{\text{emission}}+\overbrace{\sum\limits_{n}k^{\epsilon_n}_{d\to l}}^{\text{absorption}}\, ,  \nonumber \\
 k^{d,\epsilon_1}_{l\to d}&=\underbrace{\sum\limits_{m}k^{\epsilon_m}_{l\to d}}_{\text{absorption}}+\underbrace{\sum\limits_{n}k^{\epsilon_n}_{l\to d}}_{\text{emission}} 
\end{align}
where the emission and absorption rates are given by
\begin{align}
k^{\epsilon_n}_{d\to l}&=\Gamma\big[1-f(\epsilon_n)\big]n(\omega_n)\omega_n^2 \, , \nonumber \\
k^{\epsilon_m}_{d\to l}&= \Gamma\big[1-f(\epsilon_m)\big]\big[n(\omega_m)+1\big]\omega_m^2  \, , \nonumber \\
k^{\epsilon_m}_{l\to d}&=\Gamma f(\epsilon_m)n(\omega_m)\omega_m^2 \, , \nonumber \\
k^{\epsilon_n}_{l\to d}&=\Gamma f(\epsilon_n)\big[n(\omega_n)+1\big]\omega_n^2 \, .
\end{align}
Here $ n(\omega)=(e^{\omega/T_R}-1)^{-1} $ is the Planck distribution, which tells us the probability of finding a phonon with energy $ \omega $ and $ f(\epsilon)=(e^{\epsilon/T_R}+1)^{-1} $ is the Fermi-Dirac distribution, which tells us the probability of finding an occupied state at $ \epsilon $. We assume a 3D phonon density of states, thus the rates has to be multiplied by a $ \omega^2 $ term. We have absorbed all other constants from the DOS into the $ \Gamma $ introduced earlier.

The transitions between the left quantum dot and the hot left lead, i.e., the rates involving $ -(\epsilon+\epsilon_g) $, remains unchanged since we still consider this to be a large metal piece with a quasi-continuous energy-spectrum. Again, we solve the master equation in the steady state and obtain the occupation probability vector $ \mathbf{P}^{(1)} $, but now for the discrete model. With this we can find the particle currents in the channel 1 for the discrete model,
\begin{equation}
J_d^{(1)}=P^{(1)}_Rk^{d,\epsilon_1}_{d\to l} - P^{(1)}_0k^{d,\epsilon_1}_{l\to d}.
\end{equation}
Thus we can write the part of the cooling power associated with channel 1 as 
\begin{align}
\dot{Q}_R^{(1)}= 
& P^{(1)}_R\bigg(\sum\limits_{m}k^{\epsilon_m}_{d\to l}\epsilon_m+\sum\limits_{n}k^{\epsilon_n}_{d\to l}\epsilon_n\bigg) \nonumber 
\\
- & P^{(1)}_0\bigg(\sum\limits_{m}k^{\epsilon_m}_{l\to d}\epsilon_m+\sum\limits_{n}k^{\epsilon_n}_{l\to d}\epsilon_n\bigg).   
\end{align}
A similar analysis as shown here can be applied to channel 2 and provide its corresponding cooling power $ \dot{Q}_R^{(2)} $. Thus the total cooling power written as
\begin{equation}
\dot{Q}_R=\dot{Q}_R^{(1)}+\dot{Q}_R^{(2)}
\label{discrete_Q}.
\end{equation}
It should be noted that in the limit of $ T_R\to0 $ only the two levels $ \delta $ and $ \delta-\Delta $ will contribute to the total cooling power since all levels above $ \delta $ will be unoccupied and all levels below $ \delta-\Delta $ will be occupied. We can now numerically optimize 
Eq.~(\ref{discrete_Q}), with respect to the two parameters $ \epsilon $ and $ \epsilon_g $, while keeping $ T_L $ and $ T_S $ constant. Note that $ \epsilon_m $ and $ \epsilon_n $ are determined from $ \epsilon=-\epsilon_1=\epsilon_2 $ and are not free parameters. When $ \epsilon_g\gg\epsilon$, the optimal energy of the quantum dot levels $ \pm(\epsilon_g+\epsilon) $ is independent of $ \epsilon $ and therefore also independent of $ T_R $ (the only influence of $ T_R $ on those levels come via the coupling to the levels $ \pm \epsilon $). This in turn makes the optimal cooling power $ \dot{Q}_R $ approximately independent of $ \epsilon_g $.
 Hence the only free parameter for optimization is $ \epsilon(T_R) $. The plot of the optimized cooling power as a function of $ T_R $ is shown in Fig.~(\ref{fig:cooling_power}). For simplicity we set $ \delta=\Delta/2 $ and find by that the optimized cooling power as $ T_R\to 0~\mathrm{K} $ is given by
\begin{equation}
\dot{Q}_R\propto e^{-\Delta/2T_R}, \quad T_R\to 0.
\end{equation}

\subsection{Heat Capacity}
The heat capacity of a Fermi gas with temperature-independent chemical potential $\mu$  can be expressed as
\begin{equation}
C_V=\frac{dU}{dT}=\int\limits_{0}^{\infty}d\epsilon(\epsilon-\mu) D(\epsilon)\frac{\partial f(\epsilon)}{\partial T} \, .
\end{equation}
Here $ D(\epsilon) $ is the density of states (which is a constant in our case), and $ f(\epsilon) $ is the Fermi-Dirac distribution.
When going from the continuous to the discrete description we have to exchange the integral with a sum, and the continuous variable $ \epsilon $ with the discretized states $ n\Delta $.
\begin{equation}
C_V=\sum\limits_{n=0}^{\infty}\bigg(\frac{n\Delta-\mu}{T_R}\bigg)^2\frac{e^{(n\Delta-\mu)/T_R}}{(e^{(n\Delta-\mu)/T_R}+1)^2}
\label{discrete_C}
\end{equation}
This sum can easily be determined numerically, but to gain additional insight we can consider the heat capacity for a two level system. As $ T_R\to0 $ the levels $ \delta $ and $ \delta-\Delta $ will be the only relevant levels. We can write the grand canonical partition function for the two level system as
\begin{equation}
\Xi=1+e^{-\beta\delta}+e^{-\beta(\delta-\Delta)}+e^{-\beta(2\delta-\Delta)}
\end{equation}
The energy can be written as 
\begin{equation}
U=\frac{1}{\Xi}\sum\limits_{i}H_ie^{-\beta H_i}
\end{equation}
where $ H_i $ is the energy of the state $ i $. From this we can find the heat capacity from $ C_V=dU/dT $, and we find:

\begin{eqnarray} \label{heatcap}
&& C_V=  - \frac{\Delta^2 A+ \delta^2 B +\Delta \delta \,  C}{T_R^2 \left(1 + e^{-\beta\delta} +e^{-\beta(\delta-\Delta)} + e^{-\beta(2\delta-\Delta)}\right)^2}, \nonumber \\&&
A=e^{\beta(\Delta-3\delta)} + e^{\beta(\Delta-\delta)}+2e^{\beta(\Delta-2\delta)}\, ,\nonumber  \\&&
B=e^{\beta(\Delta-3\delta)} + e^{\beta(2\Delta-3\delta)} +e^{\beta(\Delta-\delta)}
+4e^{\beta(\Delta-2\delta)}+e^{-\beta\delta}, \nonumber \\&&
C=2e^{\beta(\Delta-3\delta)}+2e^{\beta(\Delta-\delta)}+4e^{\beta(\Delta-2\delta)}\, .
\end{eqnarray}
This expression is greatly simplified at  $ \delta=\Delta/2 $, i.e., a symmetric distribution of energy levels above and below $ \mu $. In this case we obtain:
\begin{equation}
C_V=\frac{dU}{dT}=2\bigg(\frac{\Delta}{2T_R}\bigg)^2\frac{e^{\Delta/2T_R}}{(e^{\Delta/2T_R}+1)^2},
\end{equation}
and with this result, and find that in the limit of $ T_R\to0 $ the heat capacity is
\begin{equation}
C_V=2\bigg(\frac{\Delta}{2T_R}\bigg)^2e^{-\Delta/2T_R}, \quad T_R\to 0.
\label{heatcap_lowT}
\end{equation}
Although this is only true for $ \delta=\Delta/2 $,  we see from the general equation for the heat capacity given in Eq.~(\ref{heatcap}) that the factor of $ T_R^{-2} $ is present for all terms, and we have found numerically that the dominating exponential terms in the optimized cooling power, 
Eq.~(\ref{discrete_Q}), and the heat capacity, Eq.~(\ref{heatcap_lowT}), always cancel each other as $ T_R\to 0 $.

\section{\label{results}Results}
\begin{figure}[b]
\centering
\includegraphics[width=1\linewidth]{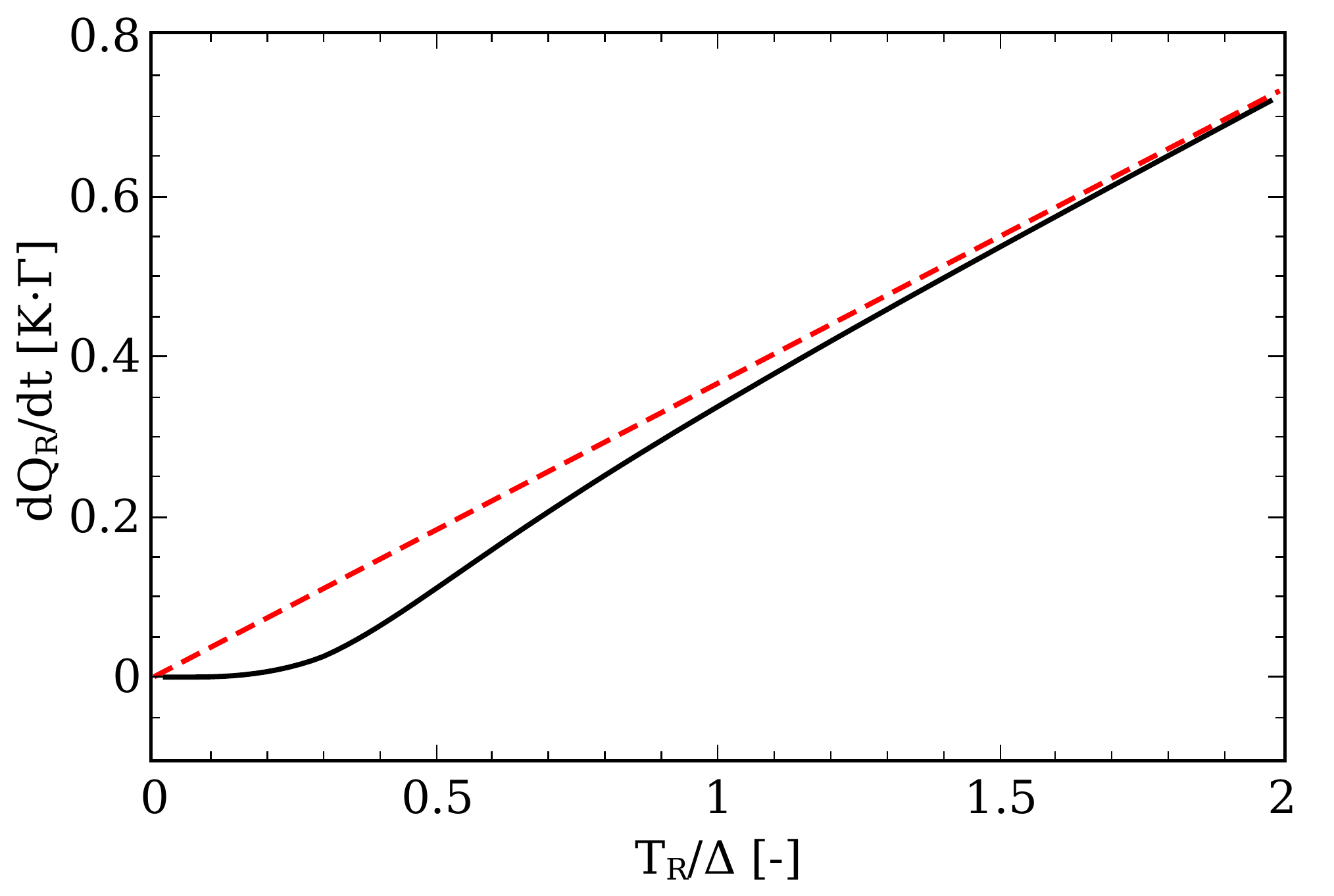}
\caption{Graph of the optimized cooling power $ \dot{Q}_R $ as a function of the dimensionless variable $ T_R/\Delta $. 
The dashed line is the result from the continuous model while the whole line is the result from the discrete model. For temperatures  $ T_R\gtrsim\Delta $ the discrete model reproduces the linear cooling power of the continuous model. However, for temperatures $ T_R\lesssim \Delta $ the cooling power changes to an exponential form. 
 Parameters used: $ \Gamma=\Gamma_s=1 $, $ T_L=20 $~K, $ T_S=6000$~K, $ \epsilon_g= 100$~K, $ \Delta=1$~K, and $ \delta=\Delta/2 $.}
\label{fig:cooling_power}
\end{figure}
\begin{figure}[ht]
\centering
\includegraphics[width=1\linewidth]{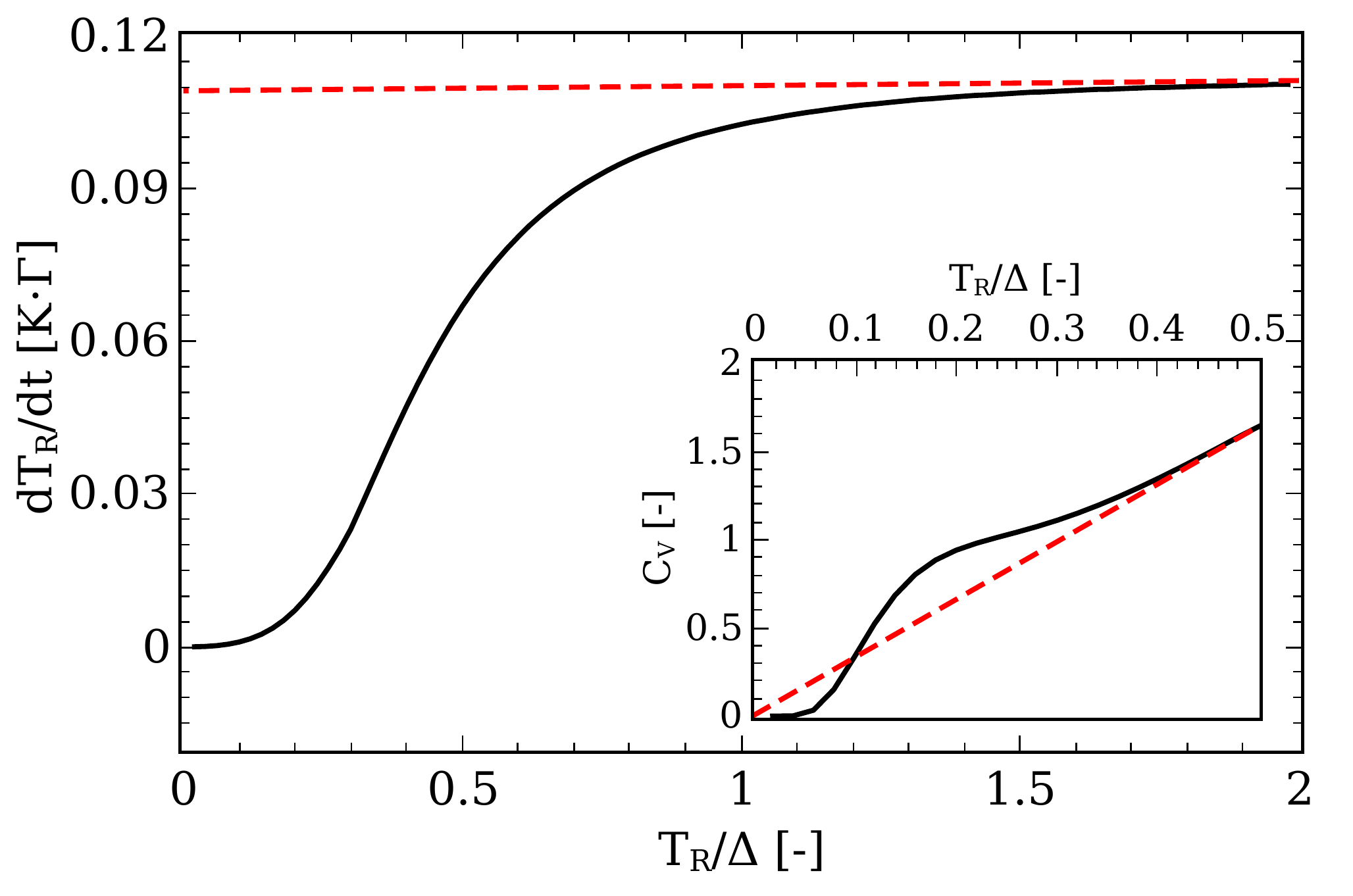}
\caption{Plot of the optimized cooling rate, $dT_R/dt $, as a function of $ T_R/\Delta $,
Again we see that the discrete model (whole line) reproduce the third law violating constant rate of temperature change of the continuous model (dashed line) for $ T_R \gtrsim \Delta $. 
The inset shows $ C_V $ as a function of the same variable. When $T_R\lesssim \Delta/2 $ the heat capacity obtains a feature similar to the Schottky-anomaly, indicating that the main contribution to the heat capacity comes from the two levels ($ \delta $ and $\delta-\Delta $) closest to $ \mu=0 $.
As a result,
 for $ T_R\lesssim\Delta $ the exponential term in $ \dot{Q}_R $ cancels the one in $ C_V $, and we are left with the $ T_R^2 $ term from the heat capacity. Parameters used: $ \Gamma=\Gamma_s=1 $, $ T_L=2$0~K, $ T_S=6000$~K, $ \epsilon_g= 100$~K, $ \Delta=1$~K, and $ \delta=\Delta/2 $.}
\label{fig:cooling_rate}
\end{figure}

We can now find the cooling rate $ dT_R/dt $ for the discrete system. In Fig.~\ref{fig:cooling_power} we have plotted the cooling power $ \dot{Q}_R $ as a function of the dimensionless variable $ T_R/\Delta $.
The whole line is the result of our numerical calculations, while the dashed line is the result form the original model~\cite{PhysRevLett.108.120603}. We see that for $ T_R \gtrsim \Delta $ the discrete model 
reproduce the results from the original model, while when $ T_R\lesssim \Delta $ the result  changes to an exponential form.

The heat capacity $ C_V $ is shown as a function of the same dimensionless variable $ T_R/\Delta $ in Fig.~\ref{fig:cooling_power}, inset. Again, it reproduces the results from the original model for $ T_R>\Delta $, but when $ T_R<\Delta/2 $ a Schottky-like feature appears, indicating that only the two levels closest to $ \mu=0 $ are participating in the dynamics.

As we discussed earlier, that if we can write the cooling rate in a form like in Eq.~(\ref{eq:unattainability}) we require that $ \alpha=\lambda-\kappa\geq1 $. In the original model with a continuous energy-spectrum in the cold metal lead, it was found that $ \alpha=0 $. Using our results from Eq.~(\ref{discrete_Q}) and Eq.~(\ref{discrete_C}), we find that for $ \delta=\Delta/2 $ the cooling rate is
\begin{equation}
\frac{dT_R}{dt}\propto \frac{\dot{Q}_R}{C_V}\propto T_R^2 \,, \quad T_R \to 0\, .
\end{equation}
We obtain $ \alpha=2 $, which implies that cooling to absolute zero is impossible in a finite amount of time, and the discrete model is thus consistent with the unattainability principle. The corresponding numerical result is shown in Fig.~\ref{fig:cooling_rate}, where we have plotted $ dT_R/dt $ as a function of $ T_R/\Delta $. Also here the result from the discrete model (whole line) reproduces the result from the original model (dashed line) for $ T_R\gtrsim \Delta $, but once $ T_R\lesssim \Delta $ it differs. For $ T_R<<\Delta $ we find by fitting the data to the function $ dT_R/dt=A~T_R^B $,  that $ dT_R/dt\propto T_R^2 $.

Although the results from Eq.~(\ref{discrete_Q}) and Eq.~(\ref{discrete_C}) are only valid for $ \delta=\Delta/2 $, we find numerically that the exponential term in $ \dot{Q}_R^{\mathrm{tot}} $ always cancels with the one in $ C_V $, and since the $ T_R^{-2} $ term in $ C_V $ is always present, the cooling rate $ dT_R/dt\propto T_R^2 $ is valid independent of choice of $ \delta $.

\section{\label{discussion}Discussion  and conclusions}
In conclusion, we have shown that our natural modification of the model proposed by Cleuren \textit{et at.} does not violate the dynamic version of the third law, and allows for the same cooling performance at temperatures $ T_R>\Delta $ as the original. This is a positive result, which tells us that the original model can be used to cool very efficiently down to the extreme limit of $ T_R\sim\Delta $, where the cooling power is quenched. Though we assumed a constant level spacing, $\Delta$, the low-temperature behavior of the cooling rate is insensitive to this assumption since at $T_R \to 0$ only the two levels closest to the chemical potential are important.

The laws of thermodynamics are so general that they should apply both to classical and quantum systems. The third law, in particular, is a theory about the properties of a system as its temperature approaches absolute zero, and at low temperatures quantum effects become important. Quantum theory predicts that confined systems have discretized energy-levels, and when the temperature $ T $ becomes comparable to the spacing between energy levels $ \Delta $, this discreteness needs to be taken into account. In \cite{PhysRevLett.108.120603} they use a continuous energy spectrum of the metal lead, disregarding the quantum discreteness. In the comments on the violation of the third law \cite{PhysRevLett.109.248901,PhysRevLett.109.248902,PhysRevLett.109.248903,PhysRevLett.112.048901}, they employ a heat capacity derived from quantum theory, and it is this mixing of classical and quantum description that leads to the breaking of the unattainability principle. If instead we use a pure classical expression for the heat capacity, which would be a constant as given by the equipartition principle, the unattainability principle would be satisfied \cite{loebl1960third}.

We have assumed that the cold metal lead equilibrates instantaneously after electron transfer. This equilibration depends on the size of the metal; larger volumes equilibrate more slowly, and the cooling would only occur in a finite volume within the metal. Not only does the size affect the equilibration time, but it also affects the spacing between the energy levels. Larger volumes results in smaller spacing and thus the system could be cooled to lower temperatures, since the discrete effects only become apparent when $  T_R\simeq \Delta $. Finding an optimal size of the system, that balances these two effects would be beneficial. Our final assumption is that the left hot lead function as a large heat bath, and have no effect on the cooling rate. A recent article \cite{Masanes2014} have shown that in a cooling process the density of states of the left heat bath affects the cooling rate of quantum refrigerators. A refined model where we take into account the properties of the left lead would give us additional insight into the nature of quantum refrigerators.
 \acknowledgments
The research leading to these results has received funding from the European Union
Seventh Framework Program (FP7/2007-2013) under grant agreement No. 308850
(INFERNOS).

%

\end{document}